\documentclass[
aps,%
12pt,%
final,%
notitlepage,%
oneside,%
onecolumn,%
nobibnotes,%
nofootinbib,%
superscriptaddress,%
showpacs,%
centertags]%
{revtex4-1}
\pdfoutput=1

\newcommand{\exclude}[1]{}

\usepackage{tikz}
\usetikzlibrary{shapes.symbols}
\usetikzlibrary{shapes.arrows}
\usetikzlibrary{positioning}
\usepackage[latin1]{inputenc}
\usepackage{listings}
\usepackage{hyperref}
\usepackage{mathrsfs}
\usepackage{amsmath}
\usepackage{colortbl}
\usepackage{slashed}

\newcommand{\eps}{\epsilon}
\newcommand{\al}{\alpha}
\newcommand{\bet}{\beta}
\newcommand{\gam}{\gamma}

\newcommand{\del}{\delta}

\begin{document}

\newcommand{\beq}{\begin{equation}}
\newcommand{\eeq}{\end{equation}}



\newcommand{\A}{\alpha}
\newcommand{\B}{\beta}
\newcommand{\T}{\theta}
\newcommand{\Ep}{\epsilon}
\newcommand{\beqn}{\begin{eqnarray}}
\newcommand{\eeqn}{\end{eqnarray}}

{}\hfill SI-HEP-2014-11

\vspace*{-8pt}

{}\hfill QFET-2014-07

\vspace*{3pt}

\title{Towards NNLO corrections in $B \rightarrow D \pi$ \vspace*{5pt}}

\renewcommand{\baselinestretch}{1.1}\normalsize

\author{\firstname{T.}~\surname{Huber}}
\email{huber@physik.uni-siegen.de}
\author{\firstname{S.}~\surname{Kr\"ankl} \vspace*{8pt}}
\email{kraenkl@physik.uni-siegen.de}

\affiliation{Theoretische Physik 1, Naturwissenschaftlich-Technische Fakult\"at,
 Universit\"at Siegen, Walter-Flex-Stra{\ss}e 3, D-57068 Siegen, Germany}


\begin{abstract}

\vspace*{8pt}

\renewcommand{\baselinestretch}{1}\normalsize

We describe the status of the calculation of the branching ratio for the decay $\bar{B}^0\rightarrow D^+ \pi^-$ to NNLO in QCD factorization. Here, we present the result for the colour-singlet hard-scattering kernel. The calculation has been performed using multi-loop techniques like Laporta's reduction to master integrals, and Mellin-Barnes representations and differential equations for evaluating the latter. 

\end{abstract}


\maketitle

\section{Introduction and Motivation}

\renewcommand{\baselinestretch}{1.1}\normalsize

Non-leptonic two-body B decays yield a broad spectrum of observables for investigating the CKM structure of the Standard Model.
The theory description of many of these observables involves matrix elements of the form $\langle f_1 f_2 | \mathcal{O}_i | B \rangle$, where the operator $\mathcal{O}_i$ describes the weak interaction of the underlying decay of the $B$ meson to some final state hadrons $f_{1,2}$.
The strong interactions of the hadronic process complicate the evaluation of these matrix elements. 
In a first approach, nowadays known as the naive factorization, the latter are expressed as products of a decay constant and a form factor, see e.g.\ \cite{Bauer:1986bm}.
Over the last two decades several methods for the description of hadronic matrix elements have been developed, the most successful ones being based on flavour symmetries (see e.g.~\cite{Jung:2009pb}) and/or factorization, like pQCD \cite{Keum:2000ph} or QCD factorization (QCDF)~\cite{Beneke:1999br,Beneke:2000ry}. The latter is a model-independent framework that systematically disentangles perturbative from non-perturbative effects in the heavy-mass limit.
Here, we apply QCDF to calculate the radiative two-loop QCD-correction to the decay $\bar{B}^0\rightarrow  D^{+}\pi^{-}$. \\
In QCDF in the heavy quark limit the amplitude for $\bar{B}^0\rightarrow  D^{+}\pi^{-}$ takes the following form~\cite{Beneke:2000ry}
\begin{align}
 \langle D^{+}\pi^{-}|\mathcal{O}_i |\bar{B}^0 \rangle = \sum_j F_j^{B\rightarrow D} (m_\pi^2) \int_0^1 du \,T_{ij}(u) \Phi_{\pi}(u)  \, . \label{bbnsfactorization}
\end{align}
$F_j^{B\rightarrow D} $ is the $B\rightarrow D$ form factor,  $\Phi_{\pi} $ the light-cone distribution amplitude (LCDA) of the pion, $T_{ij}$  the hard-scattering kernel and $u$ the momentum fraction carried by the constituents of the pion. Note that for a finite $b$ quark mass Eq.~\eqref{bbnsfactorization} receives corrections which are formally suppressed by powers of $\Lambda/m_b$, where $\Lambda$ is the typical hadronic scale. 
Weak annihilation contributions to the decay, as well as interactions involving the spectator quark of the B meson have been proven to be among such power corrections. 
 
The hard-scattering kernel contains the short distance degrees of freedom and therefore can be evaluated perturbatively as an expansion in the strong coupling $\al_s(\mu)$. It gives rise to a perturbative contribution $a_1$ which in naive factorization simply has the value $|a_1| = 1$. \\ In the case at hand in QCDF the branching ratio is given by the following expression~\cite{Beneke:2000ry}
\begin{align}
  \Gamma(\bar{B}^0 \rightarrow  D^{+} \pi^{-})= \frac{G_F^2(m_B^2-m_D^2)^2 |\vec{q}|}{16 \pi m_B^2} |V^*_{ud}	 V_{cb}|^2  |a_1(D \pi)|^2 	\, f^2_\pi \,F^2_0(m^2_\pi)  \, ,
 \end{align}
where the pion decay constant $f_\pi$ and the scalar form factor $F_0$ contain the long distance degrees of freedom.
QCDF predicts that for the set $\bar{B}^0 \rightarrow D^{(*)} L$ with  $L=\{\pi$, $\rho$, $K\}$, $a_1$ only mildly depends on the light meson $L$ which can be seen from the quasi-universality of $a_1$.
As an example the results for $a_1$ to NLO accuracy are given below where for the light meson an expansion in Gegenbauer moments up to the first moment $\al^{L}_{1}$ has been taken~\cite{Beneke:2000ry}
 \begin{align}
| a_1(\bar{B}^0 \rightarrow D L)|=&(1.055^{+0.019}_{-0.017})  -(0.013^{+0.011}_{-0.006}) \al^{L}_{1} \, , \nonumber \\
| a_1(\bar{B}^0 \rightarrow D^* L)|=&(1.054^{+0.018}_{-0.017})  -(0.015^{+0.013}_{-0.007}) \al^{L}_{1} \, . \label{a1NOL}
 \end{align}
In case of $\pi$ and $\rho$ we have $\al^{\pi(\rho)}_{1}=0$ and for the kaon $|\al^{K}_{1}|<1$ is assumed. One finds a quasi-universal value $|a_1|\simeq 1.05$.

For the decays $\bar{B}^0 \rightarrow D^{(*)} L$ with  $L=\{\pi$, $K\}$, values for $a_1$ have been recently extracted from experimental data~\cite{Fleischer:2010ca} and here, the favored central value is $|a_1|\simeq 0.95$, with errors in individual channels at the $10$-$20$\%-level.

There is still a chance that one may improve the numbers~\eqref{a1NOL} from the theoretical side. 
The NLO QCDF corrections to $a_1$ are small since they are colour-suppressed and appear alongside small Wilson coefficients. Therefore the NNLO contribution is not necessarily much smaller compared to the size of the NLO correction\footnote{Note that this does not imply a breakdown of perturbation theory, as was already pointed out for two-loop corrections to the colour-suppressed tree-amplitude in $B \to \pi\pi$ decays
~\cite{Beneke:2009ek,Bell:2007tv,Bell:2009nk}. }.
By calculating the NNLO correction we test whether the updated numbers in Eq.~\eqref{a1NOL} will still point towards a quasi-universal value $|a_1| \simeq 1.05$ or whether they shift towards the experimentally favored value.

Last but not least, the two-loop correction will most probably be the final word on the perturbative side of the factorization formula~\eqref{bbnsfactorization}. Since the perturbative series is expected to be well-behaved and also the precision on the form factor and Gegenbauer moments of the LCDA is expected to improve, this will enable us to estimate the size of power corrections by comparison to experimental data. The uncertainties of the experimental values are currently at the $5$\%-level~\cite{Amhis:2012bh},
 \begin{align}
\mathcal{B} (\bar{B}^0 \rightarrow D^+ \pi^-)=& (26.5 \pm 1.5) \cdot 10^{-4} \, , \nonumber \\
\mathcal{B} (\bar{B}^0 \rightarrow D^{*+} \pi^-)=& (26.2 \pm 1.3) \cdot 10^{-4} \, . \label{exptl}
 \end{align}

\section{Theory}

We work in the effective five-flavour theory where the top quark and the heavy gauge bosons are integrated out. The current-current operators for the $b\rightarrow c \bar{u} d$ transition are given by the effective Hamiltonian \cite{Buchalla:1995vs}
\begin{align}
 \mathcal{H}_{\rm eff}=\frac{G_F}{\surd{2}} V_{ud}^* V_{cb} \left(C_1  \mathcal{Q}_1 +C_2  \mathcal{Q}_2\right) + \text{h.c.}\, .
\end{align}
The effective four-fermion operators in the Chetyrkin-Misiak-M\"unz (CMM) basis~\cite{Chetyrkin:1997gb} read
\begin{align}
 \mathcal{Q}_1&=\bar{d} \gam_{\mu} (1-\gam_5)T^A u \,\,\,  \bar{c} \gam^{\mu} (1-\gam_5)T^A b \, ,\\
\mathcal{Q}_2&=\bar{d} \gam_{\mu} (1-\gam_5)u \, \,\, \bar{c} \gam^{\mu} (1-\gam_5) b \, . \label{Q1}
\end{align} 
$\mathcal{Q}_1$ and $\mathcal{Q}_2$ are the colour octet and singlet operator, respectively.  
They are supplemented by evanescent operators which can be found in \cite{Gorbahn:2004my,Gambino:2003zm} and which have to be taken into account to make the system of operators closed under renormalization in dimensional regularization.

\section{Matching to SCET and master formula}
The kinematics of the decay allows a matching to Soft-Collinear Effective Theory (SCET)~\cite{Bauer:2000yr,Bauer:2001yt,Beneke:2002ph,Beneke:2002ni}. We adopt the power counting $m_c/m_b \sim O(1)$ and therefore integrate out the hard fluctuations of the heavy  $b$ and $c$ field at the hard matching scale. Thus, the  QCD quark fields  $b$ and $c$ will be described by HQET fields $h_v$ and $h_{v'}$, respectively. The energetic quarks $u$ and $d$  nearly move in the same direction and therefore will be described by the same type of collinear SCET fields $\chi$. 
The matrix elements of the QCD operators $ \mathcal{Q}_i$ then can be expressed as a linear combination of a suitable basis of SCET operators\footnote{We use the same convention as in~\cite{Beneke:2009ek}.}
\begin{align}
\langle \mathcal{Q}_i \rangle = \sum_a \left[ H_{ia} \langle \mathcal{O}_a \rangle + H'_{ia} \langle \mathcal{O}'_a \rangle \right] \, ,  \label{matchingansatz}
\end{align}
where $H_{ia}$ and $ H'_{ia}$ are the bare matching coefficients. The basis of SCET operators reads\footnote{For simplicity the Wilson lines in the operators have been omitted. Since the latter are non-local on the light-cone, the correct notation would be $\bar{\chi} (t n_{-})[\dots] \chi (0)$. Therefore the coefficients $H_{ia}$ also are functions of the variable $t$ and Eq.~\eqref{matchingansatz} must be interpreted as a convolution product.}
{\allowdisplaybreaks
\begin{align}
 \mathcal{O}_1= & \bar{\chi} \frac{ \slashed{n}_{-}}{2} (1-\gam_5) \chi \hspace{2.5mm} \bar{h}_{v'} \slashed{n}_{+} (1-\gam_5) h_v \, , \nonumber\\
\mathcal{O}_2= & \bar{\chi} \frac{ \slashed{n}_{-}}{2} (1-\gam_5) \gam_{\bot}^{\al} \gam_{\bot}^{\bet} \chi \hspace{2.5mm}  \bar{h}_{v'} \slashed{n}_{+} (1-\gam_5) \gam_{\bot,\bet} \gam_{\bot,\al} h_v \,,\nonumber \\
\mathcal{O}_3= & \bar{\chi} \frac{ \slashed{n}_{-}}{2} (1-\gam_5) \gam_{\bot}^{\al} \gam_{\bot}^{\bet}\gam_{\bot}^{\gam}\gam_{\bot}^{\del} \chi \hspace{2.5mm}   \bar{h}_{v'} \slashed{n}_{+} (1-\gam_5) \gam_{\bot,\del} \gam_{\bot,\gam}\gam_{\bot,\bet} \gam_{\bot,\al} h_v \, ,  \nonumber\\
 \mathcal{O}'_1= & \bar{\chi} \frac{ \slashed{n}_{-}}{2} (1-\gam_5) \chi \hspace{2.5mm} \bar{h}_{v'}\slashed{n}_{+} (1+\gam_5) h_v \,, \nonumber \\
\mathcal{O}'_2= & \bar{\chi} \frac{ \slashed{n}_{-}}{2} (1-\gam_5) \gam_{\bot}^{\al} \gam_{\bot}^{\bet} \chi \hspace{2.5mm}  \bar{h}_{v'} \slashed{n}_{+} (1+\gam_5)\gam_{\bot,\al} \gam_{\bot,\bet}  h_v \, , \nonumber \\
\mathcal{O}'_3= & \bar{\chi} \frac{ \slashed{n}_{-}}{2} (1-\gam_5) \gam_{\bot}^{\al} \gam_{\bot}^{\bet}\gam_{\bot}^{\gam}\gam_{\bot}^{\del} \chi \hspace{2.5mm}  \bar{h}_{v'} \slashed{n}_{+} (1+\gam_5)  \gam_{\bot,\al}  \gam_{\bot,\bet}  \gam_{\bot,\gam} \gam_{\bot,\del}h_v \, . \label{Op3bar}
\end{align}
}
 Note that all operators $\mathcal{O}^{(\prime)}_a$ with index $a>1$ are evanescent.
 Their renormalization is done such that all matrix elements of evanescent operators vanish also beyond tree-level. 

Performing the matching from QCD onto SCET along the lines of~\cite{Beneke:2009ek} we arrive at the following expression for the two-loop hard-scattering kernel of the colour-singlet operator
\begin{align}
  T_{2}^{(2)}= \sum_{a=1,1'} A_{2a}^{(2),\rm{nf}} +  Z_{2j}^{(1)} A_{ja}^{(1)}+  Z_{2j}^{(2)} A_{ja}^{(0)}\, , \label{T2singlet}
\end{align}
where $j=1,2$ and the upper indices denote the loop order. For simplicity, we abbreviated the contributions from the primed operators by primed indices. The $A_{2a}^{(2),\rm{nf}}$ are the amplitudes from the set of 62 non-factorizable two-loop diagrams which can be found in~\cite{Beneke:2000ry}. The operator renormalization constants $Z_{2j}^{(\ell)}$ from the effective weak Hamiltonian are already known in the literature~\cite{Gorbahn:2004my,Gambino:2003zm}.
The quantities $A_{ja}^{(\ell)}$ denote the sum of one-loop ($\ell=1$) and tree-level ($\ell=0$) diagrams, including those of evanescent operators from $\mathcal{H}_{\rm eff}$, and including factorizable diagrams.

The analogue of the master formula~\eqref{T2singlet} for the hard-scattering kernel $T_1^{(2)}$ of the colour-octet operator $\mathcal{Q}_1$ will contain additional terms which are not present in the case of $\mathcal{Q}_2$. The reason for this is the fact that the one-loop kernel $T_2^{(1)}$ and other one-loop amplitudes of $\mathcal{Q}_2$, such as e.g.\ those stemming from the mass counterterm, do not contribute due to the vanishing of the corresponding colour factors.

\section{Computational Techniques}
The two-loop Feynman diagrams given by $ A_{ia}^{(2)\rm{nf}}$ are on-shell QCD diagrams which depend on two non-trivial scales: the momentum fraction $u$ of the constituents of the pion, and the ratio $z_c=m_c^2/m_b^2$ of the heavy quark masses. The pion mass is neglected.
We work in dimensional regularization with $D= 4-2\eps$, thus the UV and IR divergences of the two-loop diagrams appear at maximum as $1/\eps^4$ poles. A consistent treatment of $\gam_5$ in the NDR scheme with fully anticommuting $\gam_5$ is ensured by the use of the CMM basis.

The evaluation of the integrals is performed by applying commonly used multi-loop techniques like Passarino-Veltman decomposition~\cite{Passarino:1978jh} and Laporta's reduction to master integrals \cite{Chetyrkin:1981qh,Laporta:2001dd,Anastasiou:2004vj,Smirnov:2008iw}.
 
\begin{figure}[t]
\begin{minipage}{0.2\textwidth}
\vspace{5mm}
  \includegraphics[width=3.5cm]{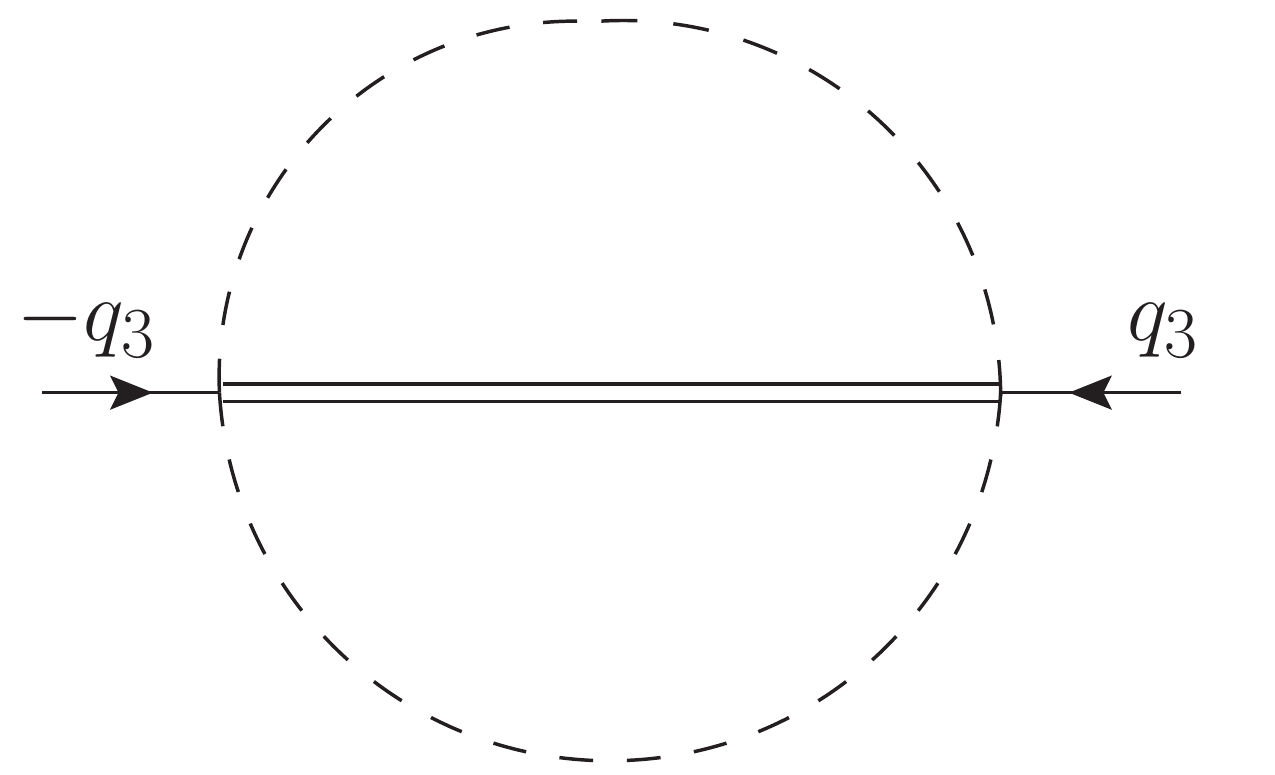} 
\end{minipage} \hspace{3mm}
\begin{minipage}{0.2\textwidth}
 \includegraphics[width=3.5cm]{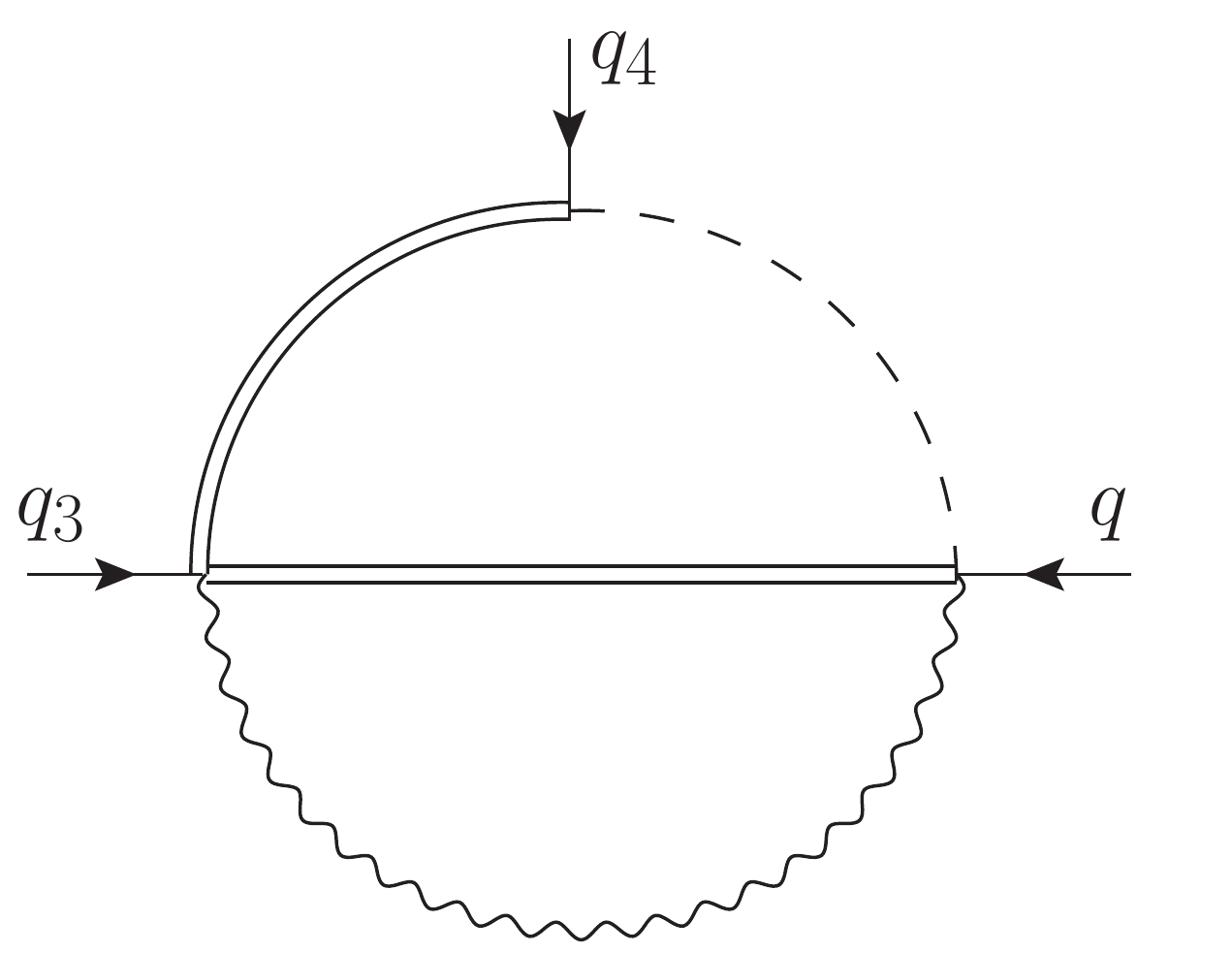} 
\end{minipage}\hspace{3mm}
\begin{minipage}{0.2\textwidth}
  \includegraphics[width=3.5cm]{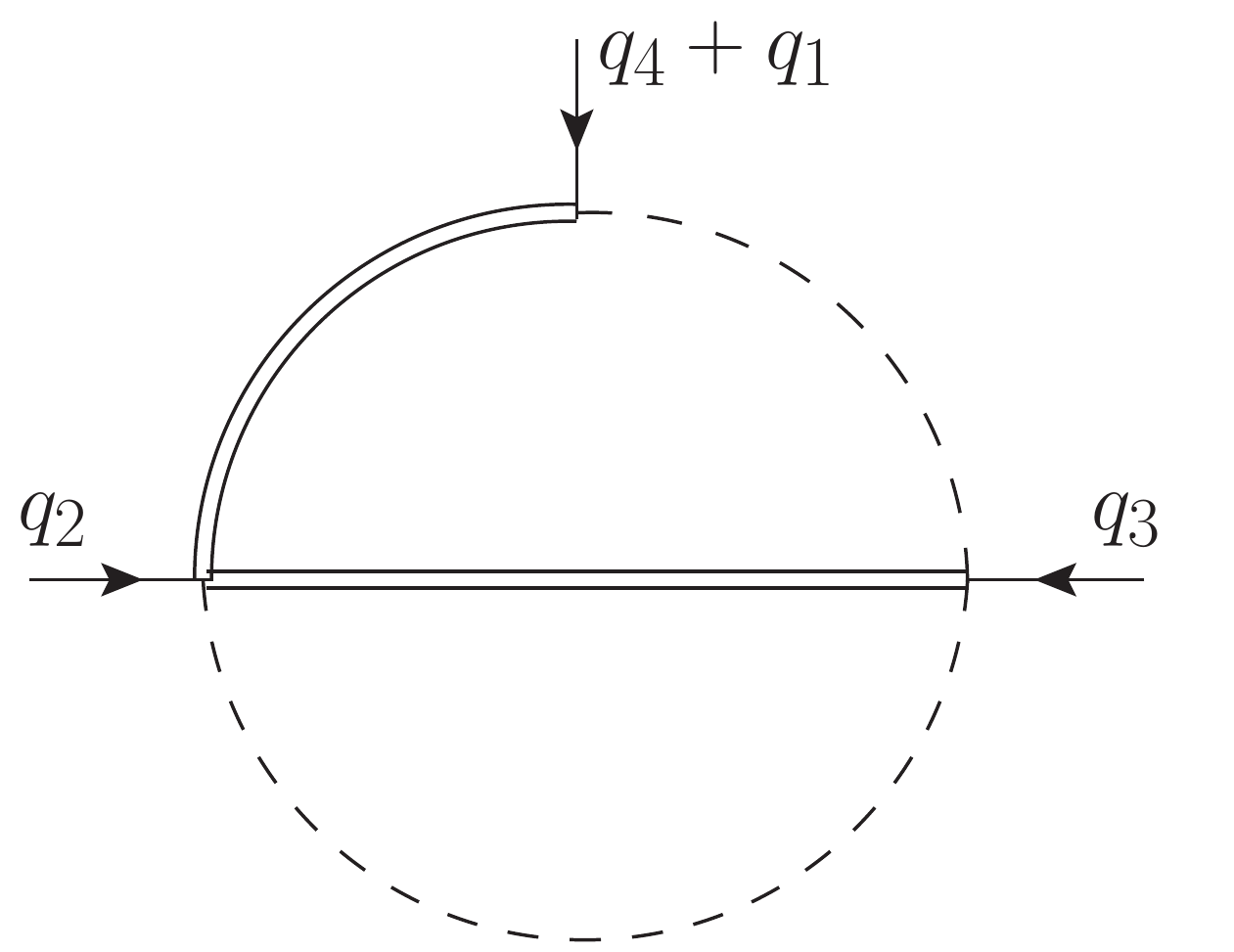} 
\end{minipage}\hspace{3mm}
\begin{minipage}{0.2\textwidth}
\vspace{5mm}
  \includegraphics[width=3.5cm]{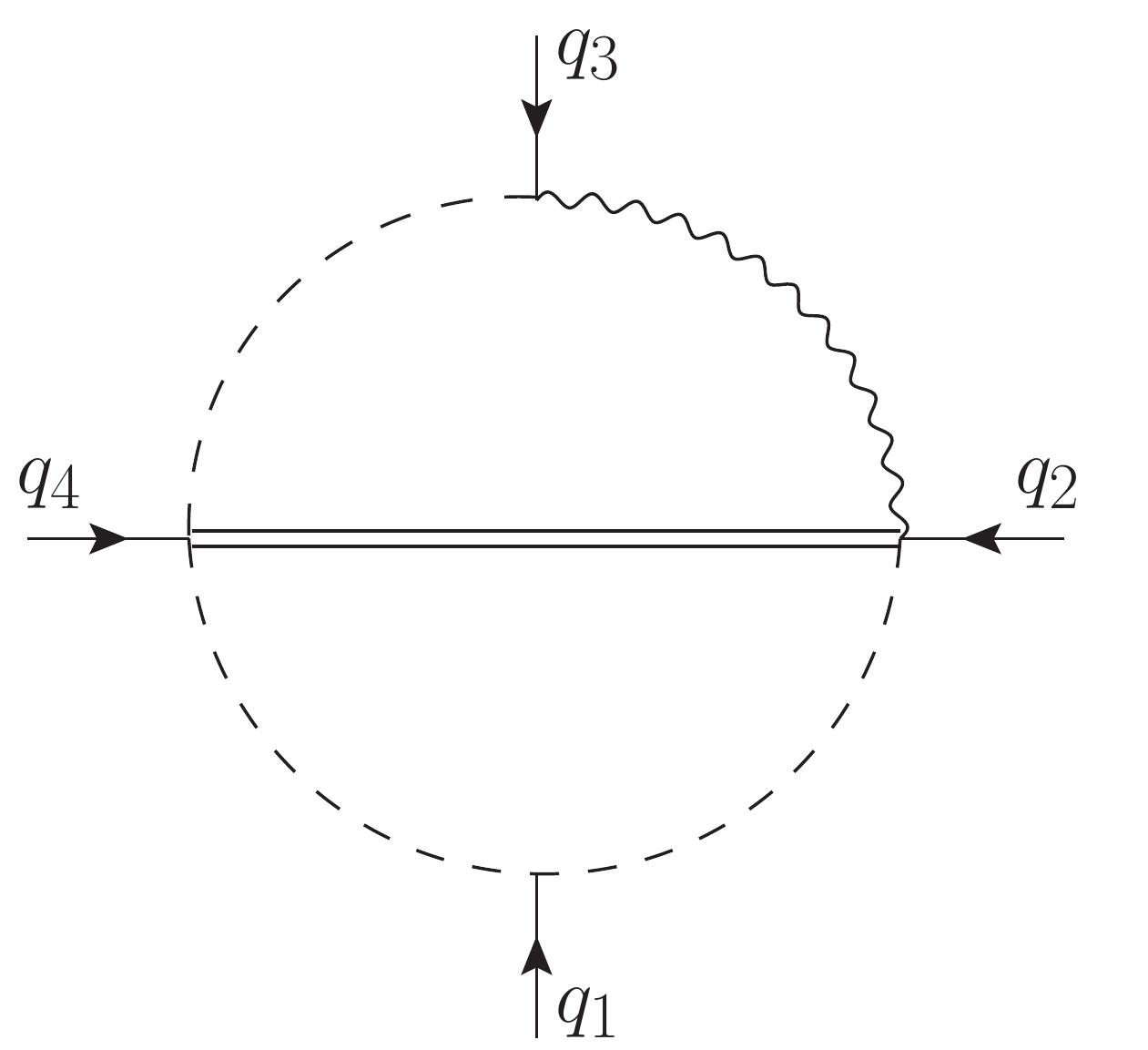} 
\end{minipage}
 \caption{Selected master integrals: The dashed, the double and the curly line represent a massless, a $b$-quark and a $c$-quark line, respectively. 
  The external momenta satisfy $q_1=q-q_2=u \; q$ and are taken to be on-shell: $q^2=0$, $q_3^2= m_c^2$, and $q_4^2= m_b^2$.}
  \label{fig:masters}
\end{figure}

We have found that there are 25 yet unknown master integrals. The topologies for a sample of them are depicted in Fig.~\ref{fig:masters}.  For their evaluation we use various techniques. 
  The results for the diagrams which are simply products of one-loop integrals can be analytically expressed in terms of hypergeometric functions. We apply the method of differential equations~\cite{Kotikov:1990kg,Gehrmann:1999as} to find analytical expressions for all master integrals that only depend on one scale, in the case at hand on $z_c$. The master integrals with three or four external legs and with both scales present are approached by introducing Mellin Barnes representations~\cite{Smirnov:1999gc,Tausk:1999vh,Czakon:2005rk,Gluza:2007rt}, therefore their result is \textquotedblleft semi-analytical\textquotedblright \, since the integrations over the Mellin-Barnes parameters only can be performed numerically. Finally we did cross-checks with the sector-decomposition program SecDec~\cite{Borowka:2012yc}.

\section{Results and Discussion}

 We obtain a numerical result for the two-loop colour-singlet hard-scattering kernel by evaluating all terms in Eq.~\eqref{T2singlet}. All poles in $\eps$ cancel at the level of $\le 6 \cdot 10^{-10}$ for about a dozen points in the $u - z_c$ plane. We therefore conclude that the expression for $T_{2}^{(2)}$ is finite in dimensional regularization, which provides a powerful check of the calculation. 
 We also seek to perform an analytical cancellation of the poles by applying the new method of differential equations for multi-scale integrals~\cite{Henn:2013pwa}.
 	
The master formula for the colour-octet hard scattering kernel $T_{1}^{(2)}$ is in preparation.
Another further step of the calculation is the convolution of the hard-scattering kernel with the light-meson LCDA to obtain a value for the perturbative quantity $a_1$. Finally, as phenomenological application we will calculate branching ratios for $B\rightarrow D^{(*)} \pi \, / \, \rho$ and confront the results with experimental data. From this comparison it will be possible to estimate the size of power corrections stemming from weak annihilation and spectator interactions.
 
\section{Acknowledgements}
S.K.\ thanks the organizers of the ITEP winter school for a pleasant and inspiring program and for the opportunity to learn cross-country skiing. 
We would like to thank Martin Beneke, Guido Bell, Bj\"orn O.\ Lange, and Thorsten Feldmann for helpful discussions, and Thorsten Feldmann for reading the manuscript. We thank A.\ Smirnov for assistance on the program FIRE. This work is supported by DFG research unit FOR 1873 ``Quark Flavour Physics and Effective Field Theories''.

\end{document}